\documentclass[nofootinbib]{revtex4}
\usepackage{tikz-feynman}
\usepackage[english]{babel}
\usepackage{array,booktabs} 
\usepackage{array} 
\usepackage{relsize}
\usepackage{calc}
\usepackage{pdflscape}
\usepackage{color}
\usepackage{graphicx}
\usepackage{amsmath}
\usepackage[nodisplayskipstretch]{setspace}

\usepackage{setspace}
\usepackage{tabularx}

\usepackage{graphics}
\usepackage{amssymb}
\usepackage[font=small,labelfont=bf]{caption}
\usepackage{graphicx}
\usepackage{epstopdf}
\usepackage{appendix}
\usepackage{soul}
\usepackage{color}
\usepackage{subcaption}
\definecolor{blizzardblue}{rgb}{0.67, 0.9, 0.93}
\definecolor{bubblegum}{rgb}{0.99, 0.76, 0.8}
\usepackage[urlcolor=blizzardblue]{hyperref}
\hypersetup{
    colorlinks=false,
    linkcolor=blue,
    filecolor=magenta,      
    urlcolor=bubblegum,
}
\begin{document}
\title{\boldmath A new viable mass region of Dark matter and Dirac neutrino mass generation in a scotogenic extension of SM}
\author{ Nilavjyoti Hazarika$^{1}$ \footnote{E-mail: nilavhazaika@gauhati.ac.in}, Kalpana Bora$^{1}$\footnote{kalpana@gauhati.ac.in}} 

\affiliation{Department Of Physics, Gauhati University, Assam, India$^{1}$}

\begin{abstract}
We propose a scotogenic extension of the Standard Model which can provide a scalar Dark Matter candidate in the new, theoretically previously unaddressed, intermediate region ($200\leq M_{DM}\leq 550$ GeV) and also generate light Dirac neutrino masses. In this framework, the standard model is extended by three gauge singlet fermions, two singlet scalar fields, and one additional scalar doublet, all of which are odd under $Z_{2} \times Z_{4}$ discrete symmetry. These additional symmetries prevent the singlet fermions from obtaining Majorana mass terms along with providing the stability to the dark matter candidate. It is known that in the case of the scalar singlet DM model, the only region which is not yet excluded is a narrow region close to the Higgs resonance {\large $m_{S} \simeq\frac{m_{h}}{2}$} - others ruled out from different experimental and theoretical bounds. In the case of the Inert doublet model, the mass region ($\sim 60$-$80$ GeV) and the high mass region (heavier than $ 550 $ GeV) are allowed. This motivates us to explore a parameter range in the intermediate-mass region $M_{W}\leq M_{DM}\leq 550 $ GeV, which we do in a scotogenic extension of SM with a scalar doublet and scalar singlets. The dark matter in our model is a mixture of singlet and doublet scalars, in freeze-out scenario. We constrain the allowed parameter space of the model using Planck bound on present dark matter relic abundance, neutrino mass, and the latest bound on spin-independent DM-nucleon scattering cross-section from XENON1T experiment. Further, we constrain the DM parameters from the indirect detection bounds arising from the global analysis of the Fermi-LAT observations of dwarf spheroidal satellite galaxies (dSphs), Higgs invisible decay and EWPT (Electroweak Precision Test) as well. We find that our model findings may provide a viable DM candidate satisfying all the constraints on DM parameters in the new, previously unexplored mass range ($200\leq M_{DM}\leq 550$ GeV). This new window for the DM candidate could be searched in furure experiments along with explanation of Dirac mass of neutrinos, since so far there is no strong evidence in support of Majorana nature of neutrino mass.
\\

\end{abstract}
\maketitle

\section{Introduction}
It is a well-established fact that the majority of the matter in the universe is made up of dark matter (DM), a non-luminous, non-baryonic form of matter whose presence is very well evident from both astrophysical and cosmological observations \cite{ParticleDataGroup:2018ovx}. The galaxy cluster observations made by Zwicky in
1930’s \cite{Zwicky:1933gu}, observations from galaxy rotation curves by Rubin in 1970’s \cite{Rubin:1970zza}, observation of the bullet cluster \cite{Clowe:2006eq} and the latest data from cosmological experiment PLANCK \cite{Planck:2018vyg} suggest that approximately 27$\%$ of the present universe is composed of
DM. This is about five times more than the ordinary visible or baryonic matter, while
the rest of it is composed of dark energy. The present abundance of DM is expressed as $\Omega_{DM}h^{2}=0.120\pm0.001$ \cite{Planck:2018vyg}, where $\Omega_{DM}$ is the DM density parameter and $h=\boldsymbol{H}/100$ $km s^{-1}Mpc^{-1}$, where $\boldsymbol{H}$ is the Hubble parameter. The Standard Model (SM) of particle physics which incorporates all the particle content of the universe is by far the most successful theory of particle physics. However, it couldn't explain the presence of dark matter and its particle nature. Deciphering the exact particle nature of DM is one of the unresolved issues in the frontiers of particle physics today. Although astrophysical and cosmological observations suggest the presence of DM, all the experiments aimed at detecting particle dark matter have so far reported null results. In the last few decades, this has motivated several beyond standard model (BSM) proposals. One of the most popular frameworks beyond SM is the so-called weakly interacting massive particle (WIMP) paradigm \cite{Kolb:1990vq}. In this framework, a DM candidate considered is typically in the electroweak scale mass range and has interaction rate similar to electroweak interactions which can give rise to the correct DM relic abundance. Further, in the Standard Model neutrino is considered to be massless. However, the evidences from Double ChooZ \cite{DoubleChooz:2011ymz} , Daya-Bay \cite{DayaBay:2012fng}, RENO \cite{RENO:2012mkc}, and T2K \cite{T2K:2011ypd} experiments hint at the neutrinos to have very small but finite masses.  For a recent update on neutrino masses, please refer to \cite{Devi:2021aaz}.

\par The singlet scalar and inert doublet models have been very widely studied in the literature. The singlet scalar model (SSM) is a simple extension of SM with a singlet scalar field $S$ added to the SM Lagrangian which is odd under a global $Z_{2}$ symmetry and all SM fields are even. This model was first considered from a cosmological point of view by Silveria and Zee \cite{Silveira:1985rk}, where the relic abundance from thermal freeze-out and direct detection cross-section for a stable real scalar gauge singlet was first calculated. Later complex scalar singlets were considered in \cite{McDonald:1993ex} and collider implication on DM-self interaction effects in \cite{Burgess:2000yq}. For recent works on scalar singlet models, one could see \cite{Yaguna:2008hd, Mambrini:2011ik} and \cite{GAMBIT:2017gge} for the current status of the scalar singlet dark matter models. In these works on the scalar singlet DM model, the only region which is not yet excluded from the observed relic abundance constraint is a narrow region close to the Higgs resonance {\large $m_{S} \simeq\frac{m_{h}}{2}$} - others ruled out from different experimental and theoretical bounds \cite{GAMBIT:2017gge}. A similar model includes a scalar doublet field being added to SM - popularly known as the Inert Doublet Model (IDM). The lightest electromagnetically neutral component of this inert doublet serves as a good DM candidate. It was first discussed as a particular symmetry breaking pattern in two Higgs Doublets Models \cite{Deshpande:1977rw}. Since then,  IDM has been widely explored in the literature, and for a recent discussions on IDM, one can refer to  \cite{Gustafsson:2012aj, Arhrib:2012ia, Goudelis:2013uca, Belanger:2015kga}. The allowed mass region where the observed relic abundance can be generated  for IDM lies in two regions - one at a low mass region below $W$ boson mass ($M_{DM} < M_{W}$) and the other at high mass region (around $ 550 $ GeV or above). The low mass region suffers strong constraints from the direct detection experiments like LUX, PandaX-II and XENON1T \cite{LUX:2016ggv,PandaX-II:2016vec,XENON:2017vdw}. These constraints reduce  the allowed DM masses in the low mass region to a very narrow mass region near {\large $m_{DM} \simeq\frac{m_{h}}{2}$}. Moreover, since so far there is no strong evidence in support of Majorana nature of neutrino mass, one can still consider models that can explain Dirac nature of light neutrino mass.

From above discussion, it is observed that the existing models cannot explain the intermediate mass range of DM candidates, and this motivated us to undertake current work. We consider scotogenic type Dirac neutrino mass model \cite{Farzan:2012sa} which can also accommodate dark matter naturally. In this BSM framework, the SM is extended by three singlet fermions which are odd under SM gauge symmetries, two singlet scalars, and an inert doublet scalar, but odd under the unbroken $Z_{2} \times Z_{4}$ symmetry. The unbroken $Z_{2} \times Z_{4}$ symmetry leads to a stable DM candidate while the $Z_{2} \times Z_{4}$ odd particles generate light neutrino masses at the one-loop level. This model belongs to the scotogenic extension of SM proposed by Ma in 2006 \cite{Ma:2006km}. The mixture of these singlet and doublet scalar fields plays the role of DM in the freeze-out scenario. This mixing among new fields opens up new annihilation and co-annihilation channels for dark matter, which contributes towards making its relic abundance close to its observed value, for the new range ($200\leq M_{DM}\leq 550$ GeV) of DM candidate.  Thus, the novel feature of this work is  that we have been able to find a new viable mass region of DM in a previously unexplored mass range (200-550 GeV) using the model discussed in detail below, and the way it can also explain the origin of light  neutrino masses (Dirac mass).  We constrain our model with reference to experimental results from XENON1T (Direct detection), Fermi-LAT (indirect detection), Higgs invisible decays from LHC, and EWPTs. If the DM is detected in this new mass window in future experiments, then our model may provide a possible viable theory for the same.

\par The paper is organised as follows. In section II, we describe the model, its particle spectrum, particle composition of DM, and origin of light neutrino masses. The theory of computing Dark matter abundance in the freeze-out scenario, DM scattering cross-section, and light neutrino masses in the model are presented in section III. Section IV contains the results of the above analysis followed by a discussion on them. We finally summarise and present a conclusion in section V.

\section{The Model}
\label{sec:headings}
As stated earlier, in this work we consider a minimal extension of the Standard Model to include Dark matter and generate neutrino masses at a one-loop level through the scotogenic framework. The SM is extended with one scalar doublet $\phi_{2}$ and one singlet $S$ with the usual Higgs doublet $\phi_{1}$. In addition, another scalar singlet field $\eta$ and three copies of vector-like fermions $N_{i}$ (i=1,2,3), apart from the SM particle content have been included. Discrete symmetries $Z_{2} \times Z_{4}$ have also been considered to forbid unwanted terms in the Lagrangian and to ensure the stability of DM. Under $Z_{2}$ symmetry, all SM-fields including $\eta$ are even (have $Z_{2}$ charge $+1$), while fields $\phi_{2}$ and $S$ and $N$ are odd (have $Z_{2}$ charge $-1$). This $Z_{2}$ symmetry could be realised naturally as a subgroup of a continuous gauge symmetry like $U(1)_{B-L}$, where $B$ is the baryon quantum number and $L$ is the lepton quantum number, with non-minimal field content \cite{Dasgupta:2014hha}. The $Z_{2}$ symmetry prevents the tree level Dirac neutrino masses term $(\overline{L}\tilde{\phi_{1}}\nu_{R})$ involving the SM
Higgs. The unbroken $Z_{2}$ symmetry
leads to a stable DM candidate and the additional discrete symmetry $Z_{4}$ prevents Majorana mass terms of singlet fermions $(M_{R}\nu_{R}\nu_{R})$.
The particle content of our model under $U(1)_{B-L}$, $Z_{2}$ and $Z_{4}$ symmetries is shown in Table I. The relevant Yukawa Lagrangian involving the lepton sector is
\begin{equation}
L_{f}=y_{l}\overline{L}\phi_{1}e_{R}+M\overline{N_{L}}N_{R}+y^{'}\overline{N_{L}}S^{\dagger}\nu_{R}+y\overline{L}\tilde{\phi_{2}}N_{R}
\end{equation}
where $\phi_{1}$ is the SM Higgs doublet and $y$, $y^{'}$s are Yukawa couplings. The neutrinos acquire a Dirac mass at one loop level as shown in the Feynman diagram in figure \ref{fig:1}.
\begin{table}[h]
\caption{Particle content of the model under $U(1)_{B-L}$, $Z_{2}$ and $Z_{4}$ }
\centering

\begin{tabular}{c c c c c c c}

\hline\hline

Particles & $U(1)_{B-L}$ & $Z_{2}$ & $Z_{4}$ & \\ [0.5ex]
\hline
$\phi_{2}$ & 0 & -1 & 1 &\\
$N_{L}$ & 1 & -1 & 1 &\\
$N_{R}$ & 1 & -1 & 1 &\\
$S$ & 0 & -1 & i &\\
$\eta$ & 0 & 1 & i &\\
$\nu_{R}$ & 1 & 1 & i &\\
$\phi_{1}$ & 1 & 1 & 1 &\\
$L$ & 1 & 1 & 1 &\\
$e_{R}$& 1 & 1 & 1 &\\
$\mu_{R}$& 1 & 1 & 1 &\\
$\tau_{R}$& 1 & 1 & 1 &\\
\hline
\end{tabular}
\label{table:1}
\end{table}

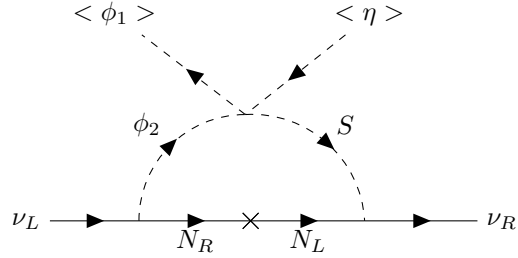
\begin{figure}[h]\label{fig:1}
\centering
\begin{tikzpicture}
\begin{feynman}[horizontal=b to d,tree layout ]
\vertex (a) {\(\nu_{L}\)};
\vertex [right=of a] (b);
\vertex [right=of b] (c);
\vertex [right=of c] (d);
\vertex [right=of d] (e) {\(\nu_{R}\)};
\vertex [above right=2cm of b] (f);
\vertex [above right=of f] (g) {\(<\eta>\)};
\vertex [above left=of f] (h) {\(<\phi_{1}>\)};

\diagram{
(a) -- [fermion] (b) -- [fermion,{edge label'=\(N_{R}\)},insertion=0.99] (c) -- [fermion,{edge label'=\(N_{L}\)}] (d)-- [fermion] (e),
(b) -- [charged scalar,{edge label=\(\phi_{2}\)}, quarter left] (f) -- [charged scalar,{edge label=\(S\)}, quarter left] (d),
(f) -- [charged scalar](h),
(g)-- [charged scalar](f),
};
\end{feynman}
\end{tikzpicture}
\caption{One-loop contribution to Dirac neutrino mass}\label{fig:1}
\end{figure}

\par The scalar potential of our model can be written as follows,
\begin{equation}
V=V_{\phi_{1}}+V_{S}+V_{\eta}+V_{\phi_{2}}+V_{int}
\end{equation}
where,
\begin{eqnarray}
\begin{array}{l}
V_{\phi_{1}}=-\mu_{1}^{2}(\phi_{1}^{\dagger}\phi_{1})+\lambda_{H}(\phi_{1}^{\dagger}\phi_{1})^{2}\\
V_{S}=\mu_{S}^{2}(S^{\dagger}S)+\lambda_{S}(S^{\dagger}S)^{2}\\
V_{\eta}=-\mu_{\eta}^{2}(\eta^{\dagger}\eta)+\lambda_{\eta}(\eta^{\dagger}\eta)^{2}\\
V_{\phi_{2}}=\mu_{2}^{2}(\phi^{\dagger}\phi)+\lambda_{\phi}(\phi^{\dagger}\phi)^{2}
\end{array}
\end{eqnarray}
and,
\begin{eqnarray}
\begin{array}{rcl}
V_{int}&=&\lambda_{3}(\phi_{1}^{\dagger}\phi_{1})(\phi_{2}^{\dagger}\phi_{2})+\lambda_{4}(\phi_{1}^{\dagger}\phi_{2})(\phi_{2}^{\dagger}\phi_{1})+\frac{\lambda_{5}}{2}[(\phi_{1}^{\dagger}\phi_{2})^{2}+ h.c]+\lambda_{6}(S^{\dagger}S)(\phi_{1}^{\dagger}\phi_{1})\\&&+\lambda_{7}(S^{\dagger}S)(\phi_{2}^{\dagger}\phi_{2})+\lambda_{8}(\eta^{\dagger}\eta)(\phi_{1}^{\dagger}\phi_{1})+\lambda_{9}(\eta^{\dagger}\eta)(\phi_{2}^{\dagger}\phi_{2})+\lambda_{10}(S^{2}\eta^{2}+h.c)\\&&+\lambda_{11}(S^{\dagger}S\eta^{\dagger}\eta)+\lambda_{12}(\eta^{\dagger}S\phi_{2}^{\dagger}\phi_{1}+h.c)
\end{array}
\end{eqnarray}
Here, $\lambda$'s and $\mu$'s are various coupling constants and mass parameters respectively. 

After spontaneous symmetry breaking, the SM Higgs $\phi_{1}$ and the inert doublet $\phi_{2}$ can be expressed as,
\begin{equation}
\phi_{1}=\begin{pmatrix} 
0\\ \frac{1}{ \sqrt{2}}(v+h)
\end{pmatrix}
\quad
,
\phi_{2}=\begin{pmatrix} 
H^{\pm}\\ \frac{1}{ \sqrt{2}}(H_{0}+iA_{0})
\end{pmatrix}
\end{equation}
The vacuum expectation value ($vev$) of the neutral component of the doublet $\phi_{1}$ is denoted by $v$.
The field $h$ corresponds to the physical SM Higgs-boson. The inert doublet consists of a neutral CP$-$even scalar $H_{0}$ with mass $m_{H_{0}}$, a pseudo-scalar $A_{0}$ with mass $m_{A_{0}}$, and a pair of charged scalars $H^{\pm}$ with mass $m_{H^{\pm}}$. The singlets $S$ and $\eta$ of the model are assumed to be complex with real part $S_{1}, \eta_{1}$ and complex part $S_{2},\eta_{2}$. The masses of the real parts are $m_{S_{1}}, m_{\eta_{1}}$ and that of the complex parts are $m_{S_{2}}, m_{\eta_{2}}$. Further, it is assumed that the real component of $\eta$ i.e. $\eta_1$ acquires a $vev$ $v_{\eta}$. In mathematical form these singlets can be written as,
\begin{eqnarray}
\begin{array}{l}
S=\dfrac{1}{\sqrt{2}}(S_{1}+iS_{2}),\\
\eta=\dfrac{1}{\sqrt{2}}[(\eta_{1}+v_{\eta})+i\eta_{2}]\\
\end{array}
\end{eqnarray}
By minimising the potential $V$ we get the masses of different physical scalars including SM Higgs and
inert particles and the singlets as,
\begin{eqnarray}
\begin{array}{l}
m_{h}^{2}=2\lambda_{H}v^{2},\\
m_{s_{1}}^{2}=\mu_{s}^{2}+\frac{\lambda_{6}}{2}v^{2}+\frac{\lambda_{11}}{2}v_{\eta}^{2}+\lambda_{10}v_{\eta}^{2},\\
m_{s_{2}}^{2}=\mu_{s}^{2}+\frac{\lambda_{6}}{2}v^{2}+\frac{\lambda_{11}}{2}v_{\eta}^{2}-\lambda_{10}v_{\eta}^2,\\
m_{\eta_{1}}^{2}=2\lambda_{\eta}v_{\eta}^{2},\\
m_{H^{\pm}}^{2}=\mu_{2}^{2}+\frac{\lambda_{3}}{2}v^{2}+\frac{\lambda_{9}}{2}v_{\eta}^{2},\\
m_{H_{0}}^{2}=\mu_{2}^{2}+\lambda_{L}v^{2}+\frac{\lambda_{9}}{2}v_{\eta}^{2},\\
m_{A_{0}}^{2}=\mu_{2}^{2}+\lambda_{L}^{'}v^{2}+\frac{\lambda_{9}}{2}v_{\eta}^{2}
\end{array}
\end{eqnarray}
where {\large $\lambda_{L}=\frac{\lambda_{3}+\lambda_{4}+\lambda_{5}}{2}$} and {\large $\lambda_{L}^{'}=\frac{\lambda_{3}+\lambda_{4}-\lambda_{5}}{2}$}. We consider a mixed state formed by the  SM Higgs and real part of singlet field $\eta$, i.e. between $h$ and $\eta_1$. The physical mass eigenstates $\zeta_1$ and $\zeta_2$ are linear combinations of $h$ and $\eta_{1}$ and can be written as,
\begin{eqnarray}
\begin{array}{l}
\zeta_{1}=h\cos \theta + \eta_{1} \sin \theta \\
\zeta_{2}=-h\sin \theta + \eta_{1} \cos \theta
\end{array}
\end{eqnarray}
The masses and mixing angle $\theta$ can be found from the diagonalisation of the mass matrix
\begin{equation}
{M_{1}}^2=\begin{pmatrix}
m_{h}^{2} & m_{\eta_{1}h}^{2}\\
    m_{\eta_{1}h}^{2} & m_{\eta_{1}}^{2}
  \end{pmatrix}=\begin{pmatrix}
2\lambda_{H}v^{2} & \lambda_{8}v v_{\eta}  \\
    \lambda_{8}v v_{\eta} & 2\lambda_{\eta}v_{\eta}^{2}
     \end{pmatrix}\\
     \end{equation}
Masses of the neutral physical scalars $\zeta_{1}$ and $\zeta_{2}$ are computed to be,
    \begin{large}
     \begin{eqnarray}
\begin{array}{l}
m_{\zeta_{1}}^{2}=\frac{m_{h}^{2} + m_{\eta_{1}}^{2}}{2} - \frac{m_{h}^{2}-m_{\eta_{1}}^{2}}{2}\sqrt{1+\tan^{2}2\theta},\\\\
m_{\zeta_{2}}^{2}=\frac{m_{h}^{2}+ m_{\eta_{1}}^{2}}{2} +\frac{m_{h}^{2}-m_{\eta_{1}}^{2}}{2}\sqrt{1+\tan^{2}2\theta}
\end{array}
     \end{eqnarray}
     \end{large}
and the mixing angle is,
\begin{equation}
\tan2\theta=\frac{2 m_{\eta_{1}h}^{2}}{(m_{h}^{2}-m_{\eta_{1}}^{2})}
\end{equation}

Similarly, the physical mass eigenstates $H_{1}$ and $H_{2}$ can be obtained as the mixture of CP-even scalars $S_{1}$ and $H_{0}$. We consider one of the mixed states $H_{2}$ as the lightest particle, so that it could serve as the DM candidate. Also, $A_{1}$ and $A_{2}$ are the physical mass eigenstates obtained from the mixing of CP-odd scalars $S_{2}$ and $A_{0}$. 
 The mass matrices are:
   \begin{large}
    \begin{equation}
     {M_{2}}^2=\begin{pmatrix}
m_{H_{0}}^{2} & m_{S_{1}H_{0}}^{2}\\
    m_{S_{1}H_{0}}^{2} & m_{S_{1}}^{2}
  \end{pmatrix}=\begin{pmatrix}
\mu_{\phi}^{2}+\frac{\lambda_{L}}{2}v^{2}+\frac{\lambda_{9}}{2}v_{\eta}^{2} & \frac{\lambda_{12}}{2}v v_{\eta} \\
    \frac{\lambda_{12}}{2}v v_{\eta} & \mu_{s}^{2}+\frac{\lambda_{6}}{2}v^{2}+\frac{\lambda_{11}}{2}v_{\eta}^{2}+\lambda_{10}v_{\eta}^{2}
     \end{pmatrix}\\
    \end{equation}
   \end{large}
   
     \begin{large}
     \begin{equation}
    {M_{3}}^2=\begin{pmatrix}
m_{A_{0}}^{2} & m_{S_{2}A_{0}}^{2}\\
    m_{S_{2}A_{0}}^{2} & m_{S_{2}}^{2}
  \end{pmatrix}=\begin{pmatrix}
\mu_{\phi}^{2}+\frac{\lambda_{L}^{'}}{2}v^{2}+\frac{\lambda_{9}}{2}v_{\eta}^{2} & \frac{\lambda_{12}}{2}v v_{\eta} \\
    \frac{\lambda_{12}}{2}v v_{\eta} & \mu_{s}^{2}+\frac{\lambda_{6}}{2}v^{2}+\frac{\lambda_{11}}{2}v_{\eta}^{2}-\lambda_{10}v_{\eta}^{2}
     \end{pmatrix}     
\end{equation}
     \end{large}

 The physical masses are obtained via the diagonalisation process of the above mass matrices respectively as,
\begin{large}
\begin{eqnarray}
\begin{array}{l}
m_{H_{1}}^{2}=\frac{m_{H_{0}}^{2} + m_{S_{1}}^{2}}{2} - \frac{m_{H_{0}}^{2}-m_{S_{1}}^{2}}{2}\sqrt{1+\tan^{2}2\alpha},\\\\
m_{H_{2}}^{2}=\frac{m_{H_{0}}^{2}+ m_{S_{1}}^{2}}{2} +\frac{m_{H_{0}}^{2}-m_{S_{1}}^{2}}{2}\sqrt{1+\tan^{2}2\alpha}
\end{array}
\end{eqnarray}
\end{large}
and
\begin{large}
\begin{eqnarray}
\begin{array}{l}
m_{A_{1}}^{2}=\frac{m_{A_{0}}^{2} + m_{S_{2}}^{2}}{2} - \frac{m_{A_{0}}^{2}-m_{S_{2}}^{2}}{2}\sqrt{1+\tan^{2}2\beta},\\\\
m_{A_{2}}^{2}=\frac{m_{A_{0}}^{2}+ m_{S_{2}}^{2}}{2} +\frac{m_{A_{0}}^{2}-m_{S_{2}}^{2}}{2}\sqrt{1+\tan^{2}2\beta}
\end{array}
\end{eqnarray}
\end{large}
The mixing angles are obtained as,
\begin{eqnarray}
\tan2\alpha=\frac{2 m_{S_{1}H_{0}}^{2}}{(m_{H_{0}}^{2}-m_{S_{1}}^{2})}\label{eq16}\\
\tan2\beta=\frac{2 m_{S_{2}A_{0}}^{2}}{(m_{A_{0}}^{2}-m_{S_{2}}^{2})}\label{eq17}
\end{eqnarray}

Various couplings of the model can be expressed in terms of the masses as,
\begin{large}

\begin{eqnarray}
\label{eq18}
\begin{array}{l}
\lambda_{H}=\frac{m_{h}^{2}}{2v^{2}},\\

\lambda_{\eta}=\frac{m_{\zeta_{1}}^{2}\sin^2\theta+m_{\zeta_{2}}^{2}\cos^2\theta}{2v_{\eta}^{2}},\\

\lambda_{8} =\frac{(m_{\zeta_{2}}^{2}-m_{\zeta_{1}}^{2})\sin\theta \cos\theta}{v v_{\eta}},\\

\lambda_{3}=\frac{2(\lambda_{L}v^{2}+ m_{H^{\pm}}^{2}-m_{H_{1}}^{2}\cos^2\alpha-m_{H_{2}}^{2}\sin^2\alpha)}{v^{2}},\\

\lambda_{4}=\frac{m_{H_{1}}^{2}\cos^2\alpha + m_{H_{2}}^{2}\sin^2\alpha+ m_{A_{1}}^{2}\cos^2\beta +m_{A_{2}}^{2}\sin^2\beta - 2 m_{H^{\pm}}^{2}}{v^{2}},\\

\lambda_{5}=\frac{m_{H_{1}}^{2}\cos^2\alpha +m_{H_{2}}^{2}\sin^2\alpha- m_{A_{1}}^{2}\cos^2\beta -m_{A_{2}}^{2}\sin^2\beta}{v^{2}},\\

\mu_{2}^{2}=m_{\zeta_{1}}^{2}\cos^2\theta + m_{\zeta_{2}}^{2}\sin^2\theta-\frac{\lambda_{L}}{2}v^{2}-\frac{\lambda_{9}}{2}v_{\eta}^{2},\\

\mu_{S}^{2}=\frac{2(m_{H_{1}}^{2}\sin^{2}\alpha+ m_{H_{2}}^{2}\cos^{2}\alpha)- \lambda_{6}v^{2}-\lambda_{11}v_{\eta}^{2}-2\lambda_{10}v_{\eta}^{2}}{2},\\

\lambda_{12} =\frac{(m_{H_{2}}^{2}-m_{H_{1}}^{2})\sin 2\alpha}{vv_{\eta}}\\

\end{array}
\end{eqnarray}

\end{large}

\subsection{Vacuum Stability}
In order that the potential in our model is bounded from below the vacuum stability requires,
\begin{equation}
 \begin{aligned}
\lambda_{H}, \lambda_{\phi}, \lambda_{S}, \lambda_{\eta} >0;\quad \lambda_{3}+2\sqrt{\lambda_{H}\lambda_{\phi}} >0\\
\lambda_{3} + \lambda_{4} - |\lambda_{5} | +  \lambda_{H} \lambda_{\phi} >0\\
 \lambda_{6}+2\sqrt{\lambda_{H}\lambda_{S}} >0; \quad \lambda_{7}+2\sqrt{\lambda_{\phi}\lambda_{S}} >0\\
\end{aligned}
\end{equation}

\subsection{Light neutrino mass}
Light neutrino masses arise at the one-loop level as shown in the Feynman diagram of figure \ref{fig:1}. To obtain Dirac neutrino mass, we consider mixing between real and imaginary components of the fields $H_{0}$ and $S_{1}$ and also between $A_{0}$ and $S_{2}$. Let, $H_{1}$ and $H_{2}$ be the mass eigenstates of $H_{0}$ and $S_{1}$ sector with a mixing angle $\alpha$ . Similarly, $A_{1}$ and $A_{2}$ be the mass eigenstates of $A_{0}$ and $S_{2}$ sector with a mixing angle $\beta$. The contribution of the real sector $(H_{0},S_{1})$ to one loop Dirac neutrino mass can then be written as \cite{Borah:2017dfn}
\begin{equation}
\label{rm}
(m_{\nu})_{R_{ij}}= \frac{\sin\alpha \cos\alpha}{32\pi^{2}}\sum_{k}y_{ik}y_{kj}^{\prime}M_{N_{K}}\left[ \frac{m_{H_{1}}^{2}}{m_{H_{1}}^{2}-M^{2}_{N_{K}}}\ln\frac{m_{H_{1}}^{2}}{M^{2}_{N_{K}}}-\frac{m_{H_{2}}^{2}}{m_{H_{2}}^{2}-M^{2}_{N_{K}}}\ln\frac{m_{H_{2}}^{2}}{M^{2}_{N_{K}}}\right] 
\end{equation}
Similarly, the contribution of the imaginary sector $(A_{0},S_{2})$ to one loop Dirac neutrino mass can be written as \cite{Borah:2017dfn}
\begin{equation}
\label{im}
(m_{\nu})_{I_{ij}}= \frac{\sin\beta \cos\beta}{32\pi^{2}}\sum_{k}y_{ik}y_{kj}^{\prime}M_{N_{K}}\left[ \frac{m_{A_{1}}^{2}}{m_{A_{1}}^{2}-M^{2}_{N_{K}}}\ln\frac{m_{A_{1}}^{2}}{M^{2}_{N_{K}}}-\frac{m_{A_{2}}^{2}}{m_{A_{2}}^{2}-M^{2}_{N_{K}}}\ln\frac{m_{A_{2}}^{2}}{M^{2}_{N_{K}}}\right] 
\end{equation}\\
where $M_{N_k}$ is the mass eigenvalue of the mass eigenstate $N_{k}$ in the internal line and the
indices $i, j = 1, 2, 3$ run over the three neutrino generations as well as three copies of $N_{i}$. $m_{H_{1}}$ and $m_{H_{2}}$ are the physical masses of the mass eigenstates $H_{1}$ and $H_{2}$, and $m_{A_{1}}$ and $m_{A_{1}}$ are the physical masses of the mass eigenstates $A_{1}$ and $A_{2}$. Thus, the total neutrino mass becomes
\begin{equation}
(m_{\nu})_{ij}=(m_{\nu})_{R_{ij}}+(m_{\nu})_{I_{ij}}
\end{equation}
where, $(m_{\nu})_{R_{ij}}$ and $(m_{\nu})_{I_{ij}}$ are defined in Eq. \ref{rm} and Eq. \ref{im} respectively.
 For this light neutrino mass to match with experimentally observed limits $(\sim 0.1)$ $eV$, we expect each of the terms of the above equation to be of sub-eV scale. Considering $M_{H_{1}} = 100$ $GeV$
and $M_{N} = 10$ $TeV$, the first term in the above expression becomes
$$(m_{\nu})^{1}_{R_{ij}}=1.46\times10^{-2}\sin2\alpha\sum _{k} y_{ik}y_{kj}$$
This could become possible naturally if  
 \begin{equation}
\sin 2\alpha y_{ik}y_{kj}^{\prime}< 10^{-8}
\end{equation}
This condition could be satisfied by suitable tuning of the Yukawa couplings $y.y'$. The mixing angles $\alpha, \beta$ depend on various masses, vev’s and relevant couplings (Eq. \ref{eq16}, \ref{eq17},\ref {eq18}). Thus, one can choose masses of various particles, and couplings of the scalar Lagrangian appropriately leading to the required mixing angles $\alpha$ and $\beta$ so that light neutrino masses agree with their observed values.

\section{Dark Matter}

As  discussed earlier, the DM candidate considered in our model is the lightest component $H_{2}$, which is a mixed singlet-doublet scalar dark matter. The thermal relic
abundance of a WIMP type DM particle $H_{2}$ is obtained by solving the Boltzmann equation \cite{Kolb:1990vq}

\begin{equation}
\frac{dn_{DM}}{dt}+3 \boldsymbol{H} n_{DM}=-<\sigma v>[n_{DM}^{2}-(n_{DM}^{eq})^{2}]
\end{equation}
where $n_{DM}$ is the number density of the dark matter particle and $n_{DM}^{eq}$ is the number density of dark matter particle when it was in thermal equilibrium. $\boldsymbol{H}$ is the Hubble rate of expansion of the Universe  and  $<\sigma v>$ is the thermally averaged annihilation cross section of the dark matter particle. One can obtain the numerical solution of the Boltzmann equation in terms of partial expansion $ <\sigma v> = a +bv^{2}$  as \cite{Scherrer:1985zt}
\begin{equation}
\Omega_{DM}h^{2}\approx\frac{1.04\times 10^9 x_{F}}{M_{Pl}\sqrt{g_{\ast}}(a+\frac{3b}{x_{F}})}
\end{equation}
where $ x_{F}=m_{DM}/T_{F}$, $T_{F}$ is the freeze-out temperature, $g_{\ast}$ is the number of relativistic degrees of freedom at the time of freeze-out.  After further simplifications, the above solution takes the form as \cite{Jungman:1995df}
\begin{equation}
\Omega_{DM}h^{2}\approx\frac{3\times 10^{-27} cm^{3}s^{-1}}{<\sigma v>}
\end{equation}

\subsection{Direct detection cross section}
The thermal averaged annihilation cross section $<\sigma v>$ is given by \cite{Gondolo:1990dk}
\begin{equation}
<\sigma v> =\frac{1}{m_{DM}^{4} T K_{2}^{2}(\frac{m_{DM}}{T})}\int \sigma (s-4 m_{DM}^{2})\surd s K_{1}(\frac{\surd s }{T})ds
\end{equation}
where $K_{i}$'s are modified Bessel functions of order $i$, $m_{DM}$ is the mass of Dark Matter particle and $T$ is the temperature of the Universe. Further, the relevant spin-independent scattering cross-section for the scalar dark matter $H_{2}$ mediated by SM Higgs can be expressed as \cite{Barbieri:2006dq}
\begin{equation}
\sigma_{SI}=\frac{\lambda_{L}^{2}f^{2}}{4\pi}\frac{\mu^{2}m_{n}^{2}}{m_{h}^{4}m_{DM}^{2}}
\end{equation}
where $m_{n}$ is the nucleon mass and {\large $\mu=\frac{m_{n}m_{DM}}{m_{n}+m_{DM}}$} is the reduced mass of dark matter and nucleon, {\large $\lambda_{L}=\frac{\lambda_{3}+\lambda_{4}+\lambda_{5}}{2}$} is the quartic DM-Higgs coupling and $f$ is the Higgs-nucleon coupling. A recent estimate of $f$ gives $f = 0.32$ \cite{Giedt:2009mr} although the full range of allowed values is $f = 0.26 -0.63 $ \cite{Mambrini:2011ik}.

\subsection{Invisible Higgs decay}
\par The DM-Higgs coupling $\lambda_{L}$ can be constrained from the latest LHC constraints on the
invisible decay width of the SM Higgs boson. The invisible Higgs decay width is related to DM-Higgs coupling $\lambda_{L}$ by \cite{ATLAS:2015ciy}.

\begin{equation}
\Gamma(h\rightarrow Invisible)=\frac{\lambda_{L}v^{2}}{64\pi m_h}\sqrt{1-4m_{DM}^{2}\diagup m_h^{2} }
\end{equation}

\subsection{Indirect detection cross section}
In addition to direct detection experiments, DM parameter space can also be
constrained using results from different indirect detection experiments like the Fermi-LAT \cite{Fermi-LAT:2015att}. These experiments search for SM particles, produced either through DM annihilations or via DM decay in the local Universe. The final photon and neutrinos, being neutral and stable can reach the indirect detection
experiments without getting affected much by intermediate regions. For WIMP type DM, these photons lie in the gamma-ray regime that can be measured at space-based telescopes like the Fermi-LAT observations of dSphs (dwarf spheroidal satellite galaxies). The observed differential gamma ray flux produced due to DM annihilations is given by
\begin{equation}
\dfrac{d\Phi}{dE}(\vartriangle\Omega)=\frac{1}{4\pi}<\sigma v>\frac{J(\vartriangle\Omega)}{2M_{DM}^{2}}\dfrac{dN}{dE}
\end{equation}
here the solid angle corresponding to the observed region of the sky is $\vartriangle\Omega$, $<\sigma v>$ is the thermally averaged DM annihilation cross section, $dN/dE$ is the average gamma ray spectrum per annihilation process. The astrophysical  factor $J$ is given by
\begin{equation}
J(\vartriangle\Omega)=\int_{\vartriangle\Omega}d\Omega'\int_{LOS}dl \rho^{2}(l,\Omega')
\end{equation}

where, $\rho$ is the DM density and LOS corresponds to line of sight. Thus, one can constrain
the DM annihilation into different final states like $\mu^{+} \mu^{-} , \tau^{+}\tau^{-}, W^{+} W^{-} , b\bar{b}$. Using the bounds on DM annihilation to these final states, we constrain the DM parameters from the global analysis of the Fermi-LAT \cite{Fermi-LAT:2015att} observations of dwarf spheroidal satellite galaxies.
\section{Results and discussion}
\label{sec:3}

In this section the results of our work and its implication have been presented, taking into account the latest constraints on dark matter relic density (Planck), spin-independent direct detection cross-section from XENON1T experiment, indirect detection bounds from Fermi-LAT, Higgs invisible decays limits from LHC, and EWPTs as well. In addition, we impose perturbativity constraint ($\lambda_{i} < 4\pi$ at all scales) on all the parameter points uniformly. We begin our analysis by imposing the Electroweak Precision Test (EWPT)\cite{Baek:2012uj} constraint in our study which provides an upper bound for the value of mixing angles as a function of the DM mass. The DM mass is allowed to vary between $5-1000$
$GeV$ and the mixing angles $\theta,\alpha,\beta$ to vary between $0$ and $1$ as shown in figure \ref{fig:2}. The couplings $\lambda_L$, $\lambda_\phi$, $\lambda_S$, $\lambda_7$, $\lambda_9$, $\lambda_{10}$, $\lambda_{11}$ with DM mass $M_{DM}$ is allowed to vary between $0$ and $1$. The black line corresponds to the EWPT (Electroweak precision test) upper bound and the uniform shaded (light orange) portion is the area ruled out by the EWPT constraints.
 \renewcommand{\thefigure}{2} 
 \begin{figure}[ht] 
\includegraphics[width=9cm, height=5cm]{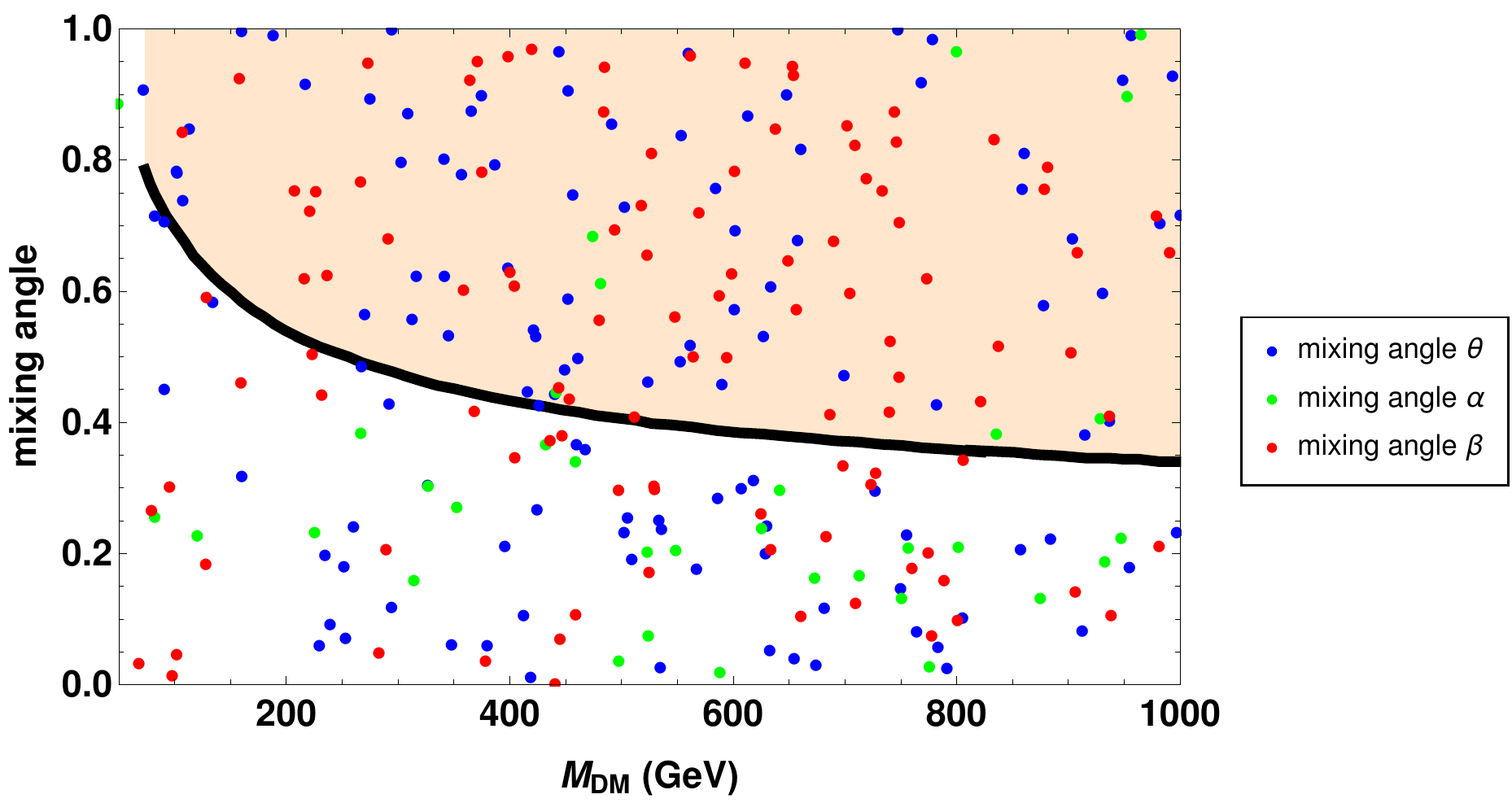}
\caption{Comparison of mixing angles $\theta,\alpha,\beta$ as a function of DM mass. The black line corresponds to the EWPT upper bound and the uniform shaded portion (light orange color) is the area ruled out by the EWPT constraints \cite{Baek:2012uj}.} \label{fig:2}
\end{figure}

We choose $H_{2}$ as the DM candidate, which is a linear combination of singlet and doublet scalars and scan the parameter space which satisfies correct relic density constraints. For computation, we have used the software package micrOMEGA 4.3.2 \cite{Belanger:2013oya} to calculate the relic abundance and spin-independent cross-section of DM. The results of our analysis are shown in figures (2-7). In figure \ref{fig:3} we have shown the variation of DM relic abundance with DM mass for our model. It is seen that there is a significant  area of parameter space which can produce correct relic density of DM in our model. Moreover, there exists a funnel-shaped region around higgs resonance $M_{DM} \approx m_{h}/2$ corresponding to the s-channel annihilation of DM into the SM fermions mediated by the Higgs boson. The observed relic abundance is satisfied for different sets of parameters.

\renewcommand{\thefigure}{3} 
 \begin{figure}[ht] 
\includegraphics[width=9cm, height=5cm]{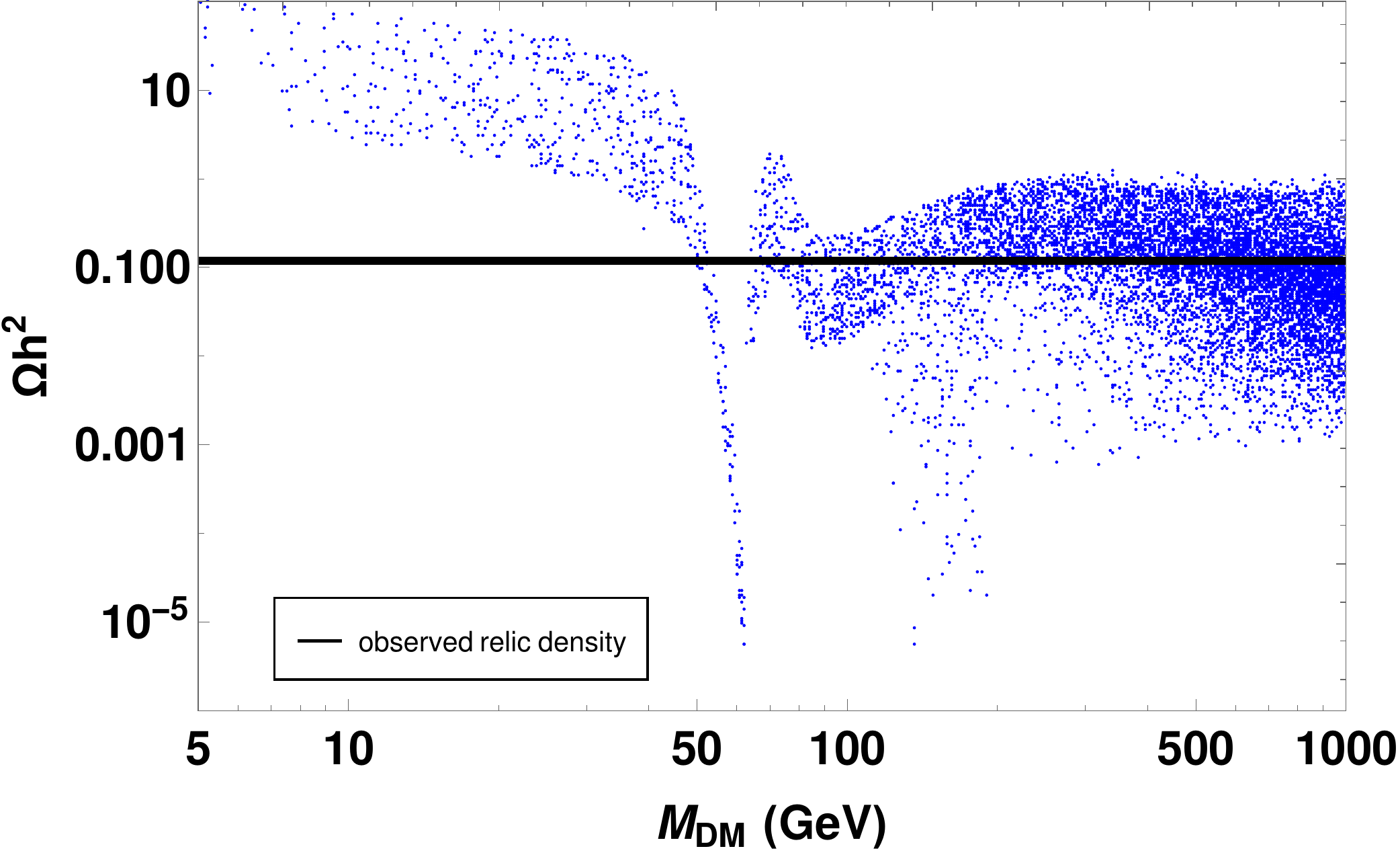}
\caption{Variation of relic density $\Omega_{DM}h^{2}$ with DM mass for parameters $\lambda_{i}= 0.01$, $\sin{\theta}=0.1,\sin{\alpha}=0.2,\sin{\beta}=0.3$ and vev $v_{\eta}=100$ GeV}. The black line corresponds to current value of DM relic density. \label{fig:3}
\end{figure}

We then scan the possible values of DM-Higgs coupling $\lambda_{L}$ for the mass squared difference $(\bigtriangleup m= m_{H_{1}}^{2}-m_{H_{2}}^{2}= 10$ $GeV$)
and show the allowed region of parameter space in $\lambda_{L}$ vs $M_{DM}$ plane from the requirement of
satisfying the correct relic abundance in figure \ref{fig:4}. The black and red exclusion lines
corresponds to XENON1T and LHC limit on Higgs invisible decay respectively. The
shaded region (light orange colored) is disallowed by LHC limits \cite{ATLAS:2015ciy} on invisible decay and direct detection of XENON1T experiment \cite{XENON:2018voc}.
The latest LHC constraint on the invisible decay width of the SM Higgs boson is applicable only for dark
matter mass $m_{DM} < \frac{m_h}{2}$ (i.e., area to the right side of red line is not allowed).

Thus, the region of parameter space to the right of red line and below black line satisfy both LHC and XENON1T limits. Hence,  one can justify to choose lower vales of coupling $\lambda_L \le 0.1$ in our analysis. The scalar DM in the model can give rise to DM spin-independent scattering cross section with nucleons that are tightly constrained by the recent bounds from direct detection experiments like LUX, PandaX-II, and XENON1T \cite{LUX:2016ggv,PandaX-II:2016vec,XENON:2017vdw}. It would be interesting to investigate the effect of variation of various coupling constants. Hence, in figure \ref{fig:5}, dependence of spin-independent DM-nucleon cross-section $\sigma_{SI}$ ($cm^{2}$) on coupling $\lambda_{L}$ points, allowed from correct relic density constraint is shown. The results are compared with spin-independent DM-nucleon cross-section bounds from latest XENON1T experiment. From the spin independent direct detection cross section plot of figure \ref{fig:5}, we observe that many points in the parameter space, satisfying correct relic density constraint \cite{Planck:2018vyg} also lie in the allowed region of XENONIT \cite{XENON:2018voc} experimental curve (below black curve).  Thus, it can be stated that the DM candidate with mass $M_{DM}\geq 200$ GeV and with small coupling $\lambda_L$ satisfies the experimental constraints of both the correct relic density and the DM-nucleon scattering cross section.
\renewcommand{\thefigure}{4} 
 \begin{figure}[ht] 
\includegraphics[width=8cm, height=5cm]{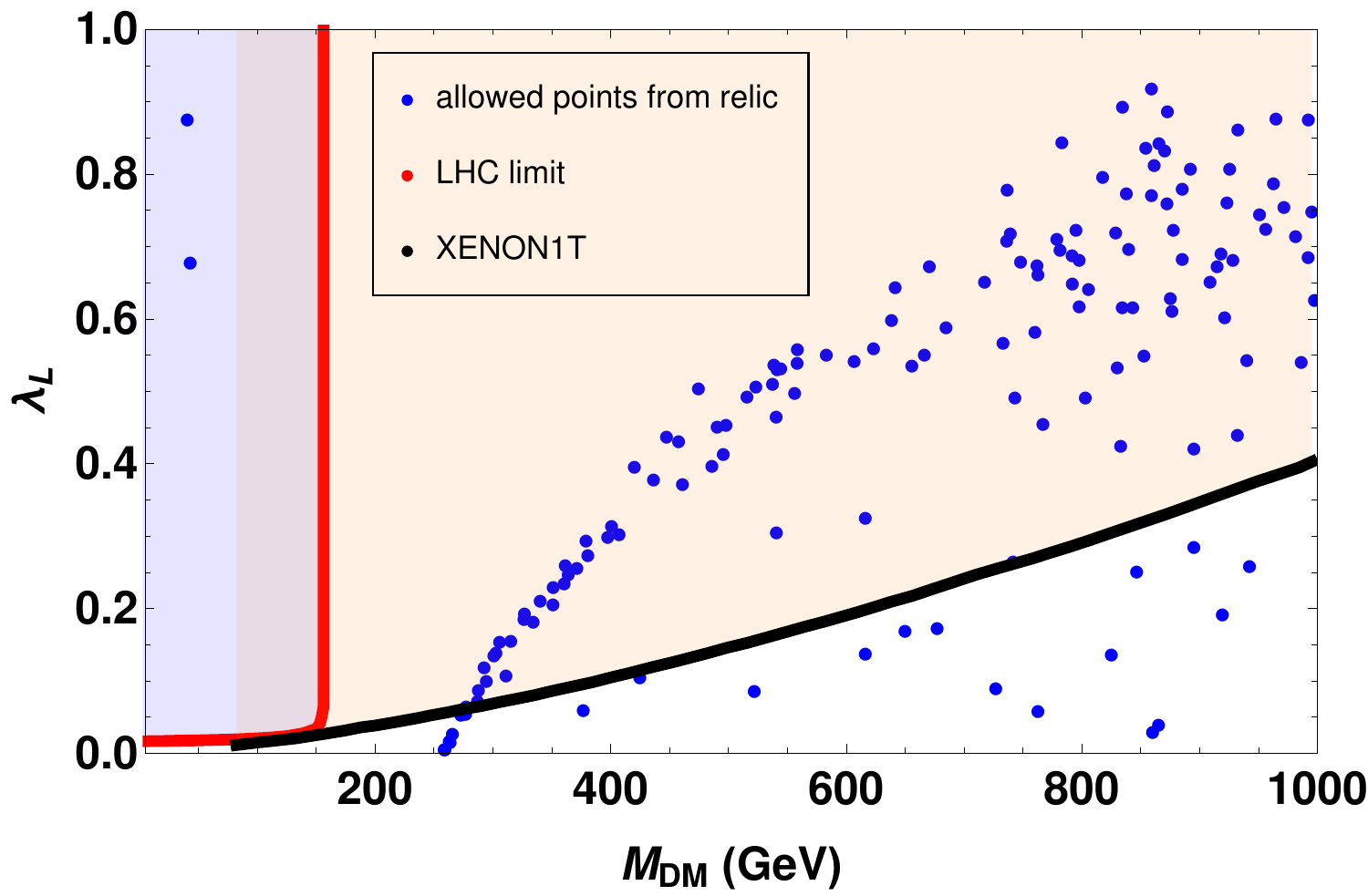}
\caption{Relic density allowed points in $\lambda_{L}$ vs $M_{DM}$ plane. The blues points are allowed from correct relic density constraints. The black and red exclusion lines correspond to XENON1T and LHC limit on Higgs invisible decay respectively (area to the right of red line and below black curve are allowed by current experimental limits).} \label{fig:4}
\end{figure}

\renewcommand{\thefigure}{5} 
 \begin{figure}[ht] 
\includegraphics[width=8cm, height=5cm]{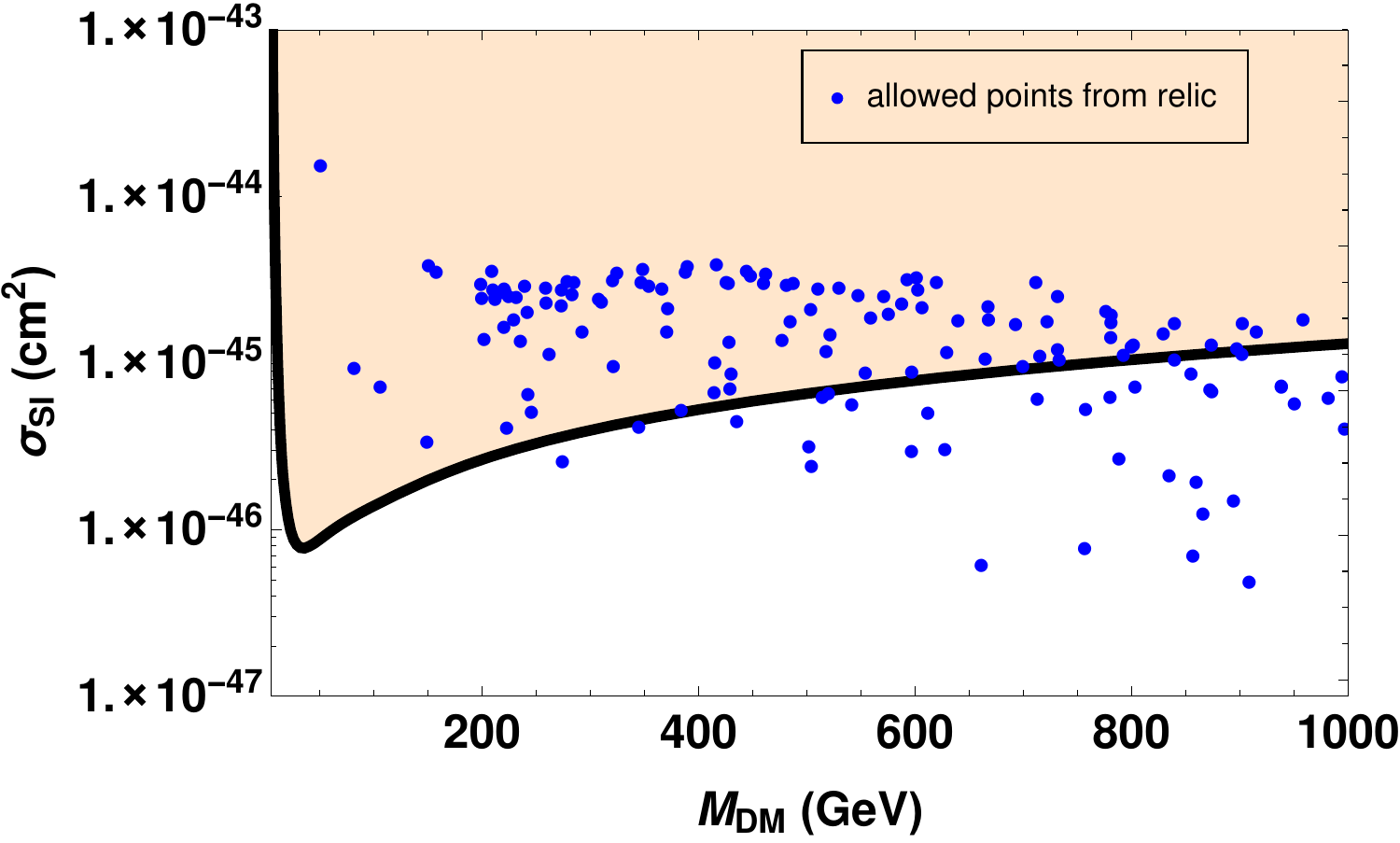}
\caption{Spin-independent DM-nucleon cross-section $\sigma_{SI}$ ($cm^{2}$) for points allowed from correct relic density constraint. The black curve is the exclusion curve from XENON1T experiment. The blue colour points are allowed from correct relic density constraint. The coupling $\lambda_L$ is varied in this plot.} \label{fig:5}
\end{figure}
Next, we study how the variation of other coupling constants affect the analysis, and in figure \ref{fig:6}, the variation of spin independent DM- nucleon cross section $\sigma_{SI}$ with DM mass $M_{DM}$ is presented with the constraint of satisfying the correct relic abundance for different  couplings $\lambda_7$, $\lambda_9$, $\lambda_{10}$,  $\lambda_\phi$, $\lambda_S$. Thus, by varying the couplings, the correct relic density could be obtained in the allowed region from spin independent DM- nucleon cross section $\sigma_{SI}$ constraint. Other parameters considered here are set to values as,
\begin{equation}
\sin{\theta}=0.1, \quad \sin{\alpha}=0.2, \quad \sin{\beta}=0.3, \quad v_{\eta}=100\quad GeV. 
\end{equation}
From our results in figure \ref{fig:6}, it is clear that our model can predict DM candidate satisfying the relic and cross section constraints, for a large range of DM mass, and this validates our model.

\renewcommand{\thefigure}{6}
 \begin{figure}[ht] 
\includegraphics[width=10cm, height=5cm]{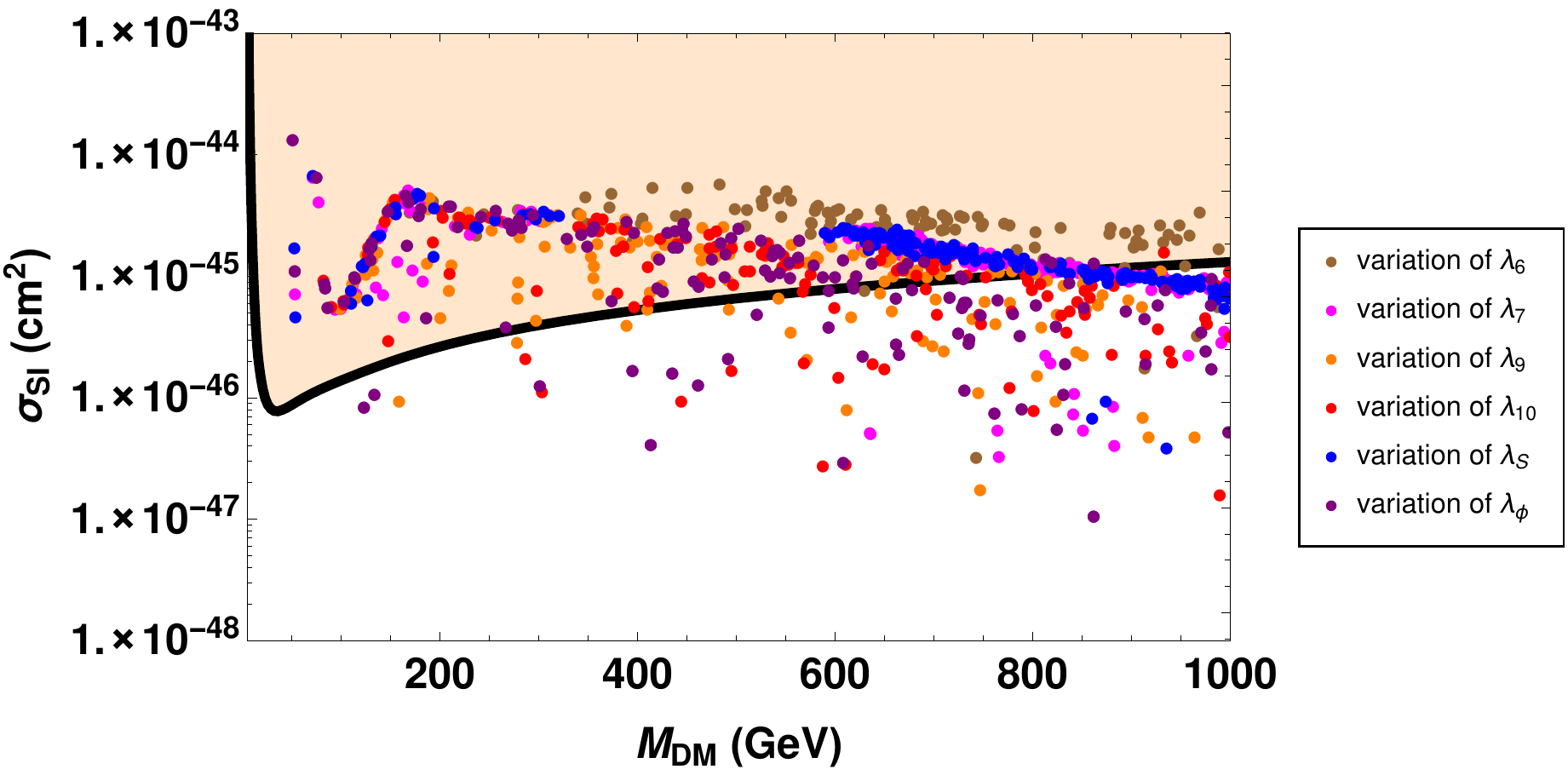}
\caption{ Spin independent DM- nucleon cross section $\sigma_{SI}$ ($cm^{2}$) for points satisfying correct relic density. The black curve is the exclusion curve from XENON1T experiment. The region below black curve is allowed from experimental constraints. } \label{fig:6}
\end{figure}

As the DM in our model is a mixture of singlet and doublet scalars, hence in addition to constraints from direct detection experiments, the DM parameter space can also be
probed in different indirect detection experiments.  We constrain the DM parameters from the indirect detection bounds arising from the global analysis of the Fermi-LAT \cite{Fermi-LAT:2015att} observations of dwarf spheroidal satellite galaxies (dSphs). In figure \ref{fig:7} we have shown DM annihilation cross section into $\tau^{+}\tau^{-}$ , $W^{+}W^{-}$ final states and compared the results with the latest
indirect detection bounds of Fermi-LAT \cite{Fermi-LAT:2015att}. In figure \ref{fig:7},  the regions below red curve is allowed, and it is seen that the previously ruled out DM mass range ($200\leq M_{DM}\leq 550$ GeV) could generate correct relic abundance in our model, and is also allowed from the DM annihilations bounds. In low mass region the s-channel dark matter annihilation
into the SM fermions through Higgs mediation dominates over other channels. The annihilation cross section of DM also has additional contributions from co-annihilations between dark matter $H_{2}$ and
heavier components $H_{1}$ \cite{Edsjo:1997bg, Bell:2013wua}. We have noted that the main processes which contribute to the relic abundance of DM
are: $H_{2} H_{2}\rightarrow f \bar{f},h h, W^{+} W^{-}, Z Z, h \zeta_{2}$; $H_{1} H_{2}\rightarrow f \bar{f}, h \zeta_{2}$ (for $\zeta_{2}$, please see Eq. (8)), and  $f$ refers to SM fermions. These new annihilation and co-annihilation channels contribute in generating the correct relic abundance of dark matter in its new  intermediate mass range. Hence, the mixed singlet-doublet scalar DM candidate $H_{2}$ in this model can very well satisfy both the relic abundance and direct detection cross section constraints for the new mass range $200-550$ GeV, and this new viable mass region is the novelty of this work.
\renewcommand{\thefigure}{7} 
 \begin{figure}[ht] 
\begin{subfigure}{0.48\textwidth}
\includegraphics[width=8cm, height=5cm]{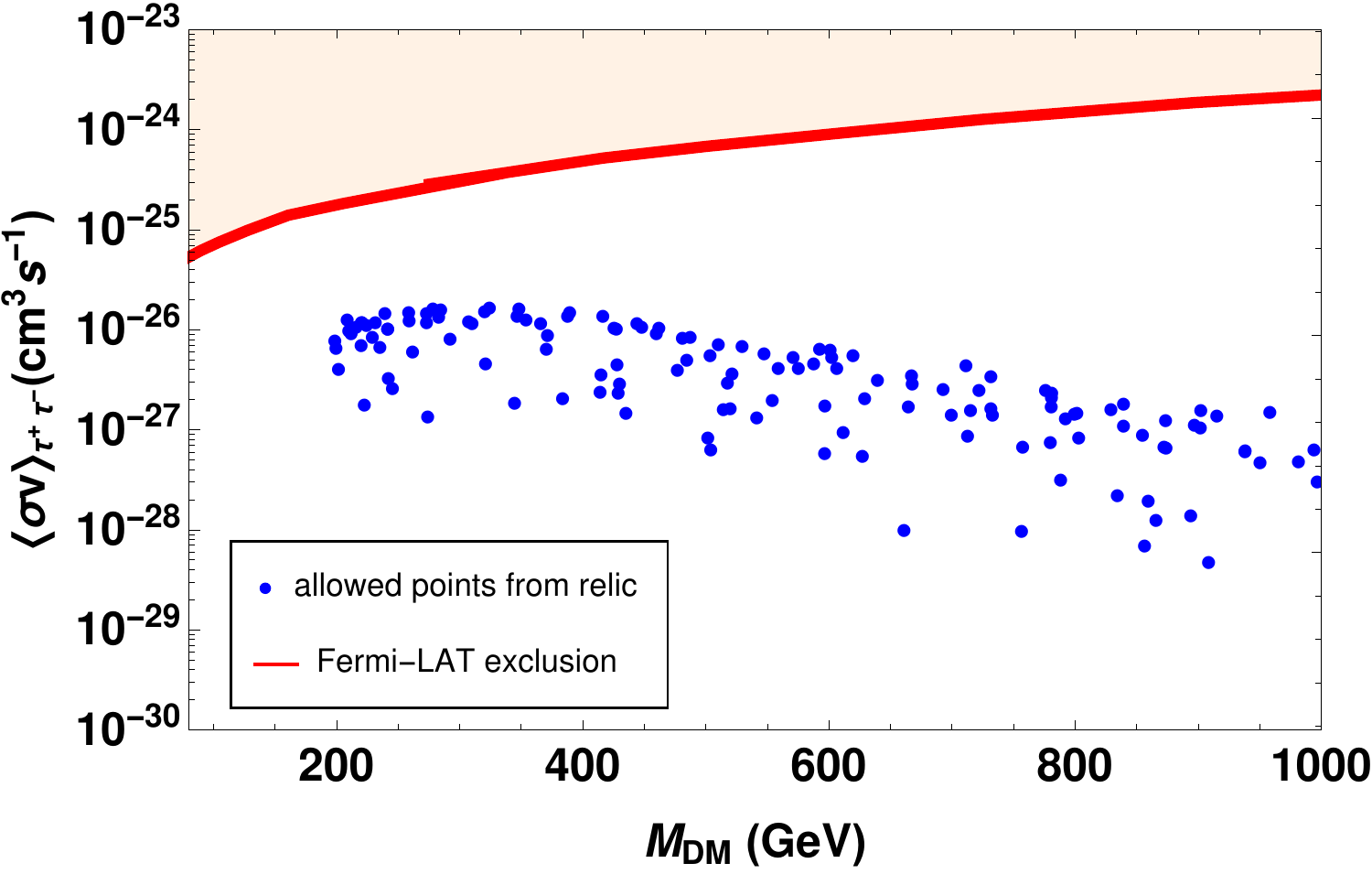}
\caption{DM annihilations into $\tau^{+}\tau^{-}$} \label{fig:7a}
\end{subfigure}\hspace*{\fill}
\begin{subfigure}{0.48\textwidth}
\includegraphics[width=8cm, height=5cm]{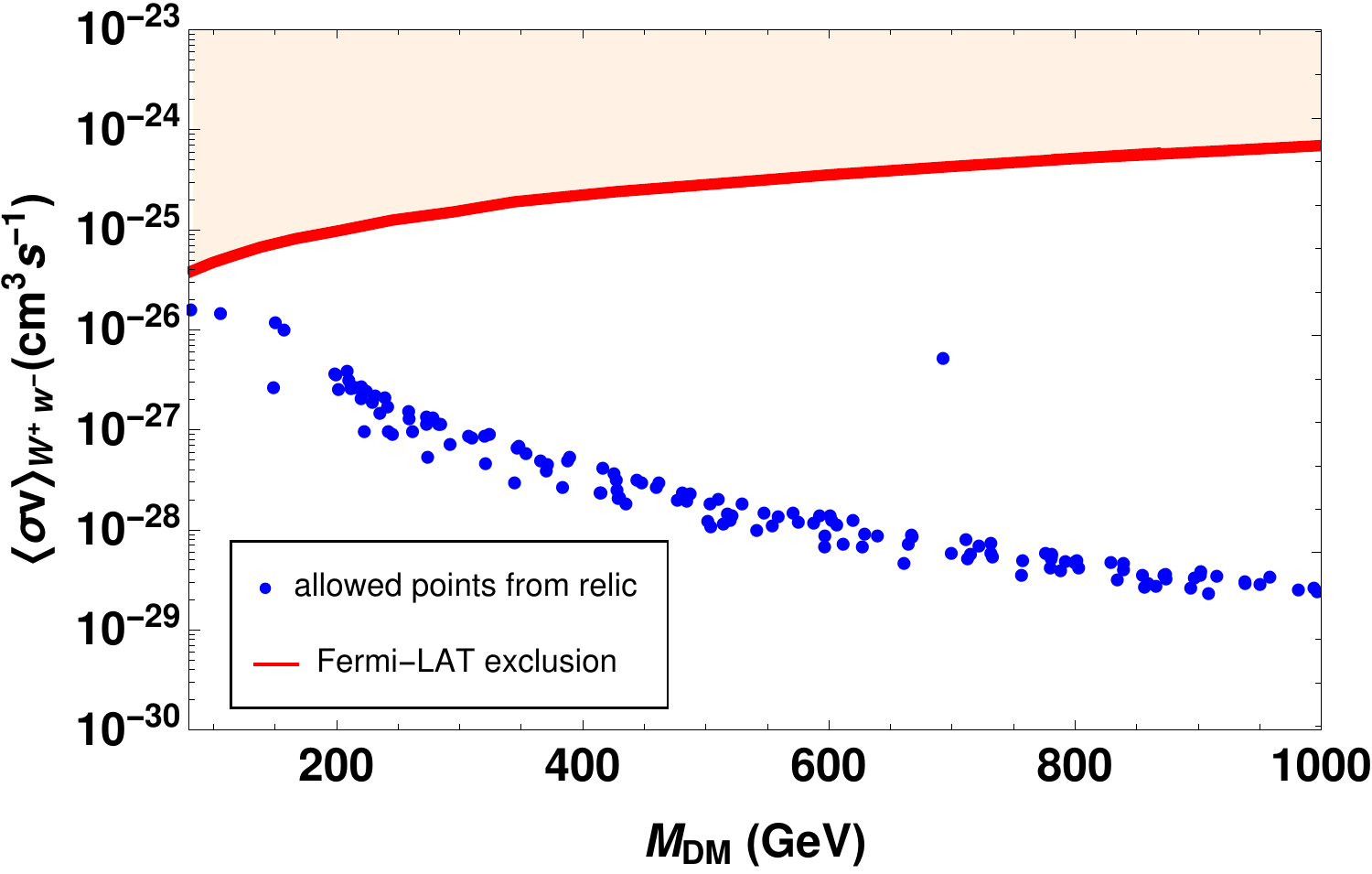}
\caption{DM annihilations into $W^{+}W^{-}$ } \label{fig:7b}
\end{subfigure}
\caption{DM annihilations into $\tau^{+}\tau^{-}$(left), $W^{+}W^{-}$ (right) compared against the latest
indirect detection bounds of Fermi-LAT. The regions below red line are allowed.}\label{fig:7}
\end{figure}

\par Thus, in this model, it has been possible to obtain the feasible DM candidate in previously unexplained mass range $200\leq M_{DM}\leq 550$, which makes the model very interesting, and is the novelty of this work. Existence of this medium mass range of the DM has been made possible due to opening of new annihilation channels of DM into particles in new mass ranges (as new scalars with varying couplings and mixings are present in the model), which brings the DM relic density in the allowed range. Also, as is clear from Eqs. (19-22), it can also explain the sub-eV Dirac neutrinos. 

\section{Conclusions}
\label{sec:5}

In this work we proposed a dark matter model with the scotogenic extension of SM, where the DM $H_{2}$ is a mixture of singlet-doublet scalars. We studied the possibility of generating scalar dark matter in the previously disallowed DM mass window of $M_{W}\leq M_{DM}\leq 550 $ GeV along with generating small Dirac neutrino mass (since Majorana nature of neutrinos is not yet confirmed). This region was disallowed in the Inert Doublet model from various experimental bounds as mentioned earlier. We impose constraints on the couplings from perturbativity ($\lambda_{i} < 4\pi$ at all scales) considerations on all the points. Also the mixing angles $\theta,\alpha,\beta$ taken are allowed by the EWPT constraints. Further, the DM-Higgs coupling is chosen such that it satisfies the limits from XENON1T (DD) and LHC limit on Higgs invisible decay. We scanned the parameter space of our model which satisfies the latest relic density and spin-independent DM-nucleon scattering cross-section constraints. It is observed from our results in the figures (2-7) that DM mass in the new range, i.e. $200\leq M_{DM}\leq 550$ GeV which was previously not viable, satisfies the relic density bound (from Planck experiment) as well as the spin-independent cross section bound from XENON1T experiment. Further, the DM parameters very well satisfy the constraints from the indirect detection bounds arising from the global analysis of the Fermi-LAT observations of dSphs. In our opinion, feasibility of this new mass range of the DM has been made possible due to opening of new annihilation and co-annihilation channels of DM in the new mass ranges. Moreover, it is observed that by choosing different mass terms and couplings of the scalar Lagrangian appropriately one can obtain the Dirac neutrino mass in the sub-eV scale as well. It may be noted that Majorana nature of neutrinos has not been established  experimentally so far. Hence, this model can explain the origin of DM in this new intermediate-mass window if detected in future experiments, which was previously unaddressed in currently available scalar DM models. 

\section{Acknowledgment} 

We acknowledge the RUSA and FIST grants of Govt. of India for support in upgrading computer laboratory of the Physics Department of Gauhati University, where this work was completed. We also thank Dr. Debasish Borah of IIT Guwahati for his valuable discussions and suggestions, during initial stages of the work.

\section{Declarations}
\textbf{Conflicts of interests:} The authors declare no potential conflict of interests.

\end{document}